\documentclass[10pt]{article}

\usepackage{graphicx}
\usepackage{cite}
\begin{document}
\begin{center}
{\large \bf
Studies of the $\eta$ meson \\
with  WASA at COSY and KLOE-2 at DA$\Phi$NE
}

\vspace{4mm} 
 
Pawe{\l} Moskal  for and on behalf of the KLOE-2 and WASA-at-COSY Collaborations
 
\vspace{2mm}
Institute of Physics, Jagiellonian University, PL-30-059 Cracow, Poland

Institut f{\"u}r Kernphysik, Forschungszentrum J\"{u}lich, D-52425 J\"ulich, Germany
\end{center}

\hspace{-0.7cm} {\bf Abstract}

One of the basic motivations of the KLOE-2  and WASA-at-COSY experiments
is the test of fundamental symmetries and the search for phenomena beyond the Standard Model
in the hadronic and leptonic decays of ground-state mesons and in particular 
in decays of the $\eta$ meson.
At COSY these mesons are  produced in collisions of proton or deuteron beam 
with hydrogen or deuterium pellet target,
and at DA$\Phi$NE $\eta$ mesons originate from radiative decays
of $\phi$ meson  or from the fusion of virtual gamma quanta
exchanged between colliding electrons and positrons. 
This contribution includes brief description of experimental techniques used by KLOE-2 and WASA-at-COSY
as well as some of physics aspects motivating investigations 
of production and decays of $\eta$  mesons.



\section{Introduction}
    Investigations of $\eta$ mesons production and decay  constitute part of the 
    experimental programmes realized with WASA~at~COSY and KLOE-2~at~DA$\Phi$NE experimental facilities.
    The cooler synchrotron COSY enables production of mesons in the hadronic interactions and 
    the electron-positron collider DA$\Phi$NE facilitates production of mesons 
    in the electromagnetic interactions.
    Thus realization of the two experimental programmes provides 
    complementary results obtained with not only 
    different detectors but also with utterly different physical  and instrumental backgrounds.

    The $\eta$ and $\eta^{\prime}$  mesons possess many interesting features making them particularly
    suitable for investigations of e.g. 
    (i) discrete symmetries, 
    (ii) anomalies of Quantum Chromo Dynamics (QCD) and
    (iii) hadronic interactions.
    Studies of properties of these mesons open many possibilities
    for searching of new kind of matter as eg.  
    mesic-nuclei, dark matter bosons, gluonium content in mesons, and give a chance to observe processes
    which are not described in the framework of the Standard Model (SM).  
    A comprehensive and detailed description of physics motivations 
    for studies of $\eta$ and $\eta^{\prime}$ 
    mesons production and decays  is included in physics reports of KLOE-2~\cite{kloe2amelino} 
    and WASA-at-COSY~\cite{wasaatcosyproposal} experiments. 
    Therefore in this contribution we will only briefly 
    and in general terms discuss some of physics aspects of these experiments, 
    and for the detailes the interested reader
    is referred to the abovementioned articles.
 
    In the KLOE experimental campaign completed in the year 2005 about 10$^8$  
    events with the $\eta$ mesons have been collected and 
    this ensemble will be increased to about 10$^9$ 
    within next three/four years of the KLOE-2 running.  
    Similarly WASA-at-COSY has collected so far about 
    3 x 10$^8$ events with the $\eta$ meson and expects to increase this sample 
    up to about $10^9$ events within next four years of running.

\section{KLOE-2 at DA$\Phi$NE}
\begin{figure}[h]
       \includegraphics[width=5.5cm]{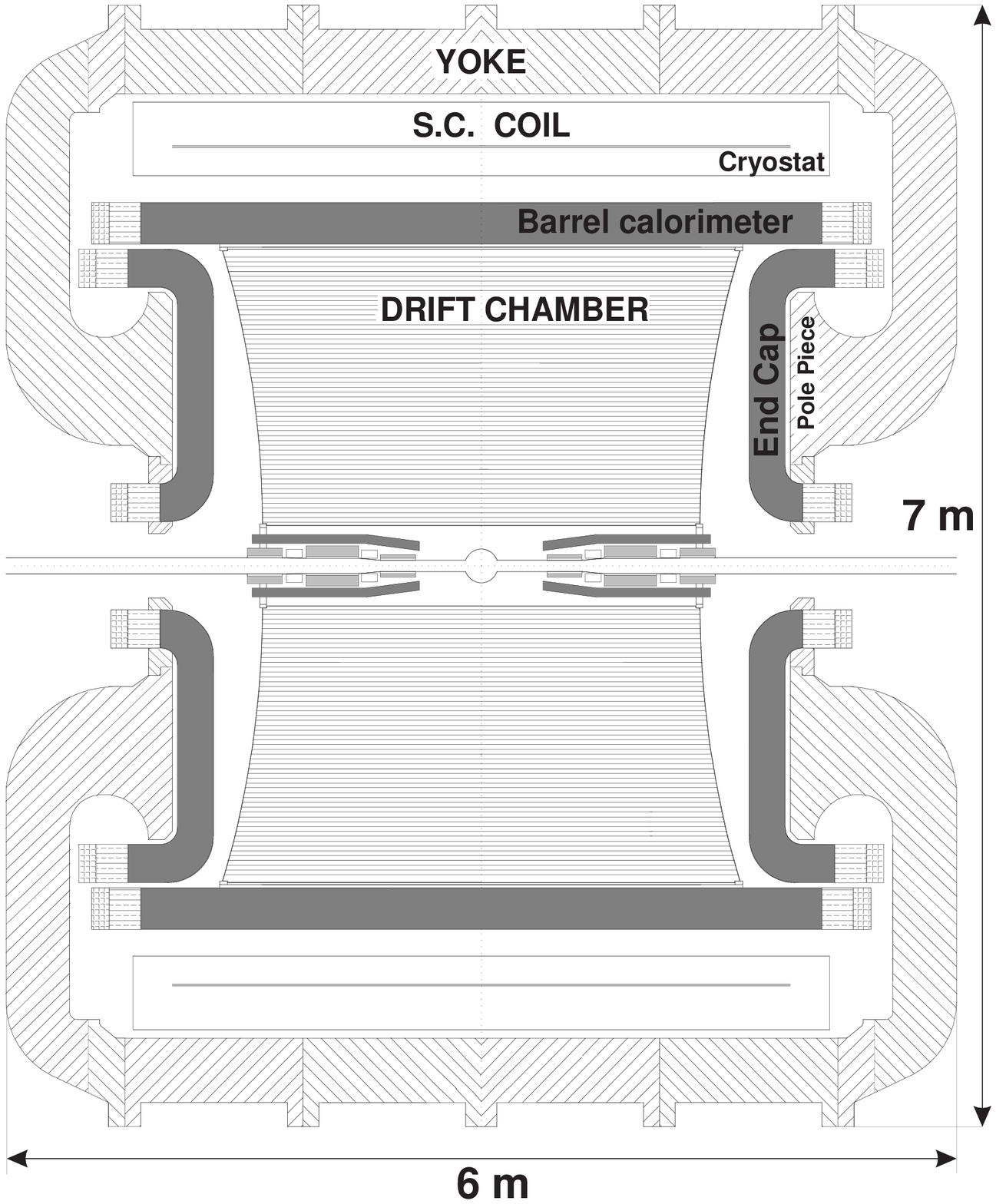}\hspace{0cm}
       \includegraphics[width=7.5cm,height=6.4cm]{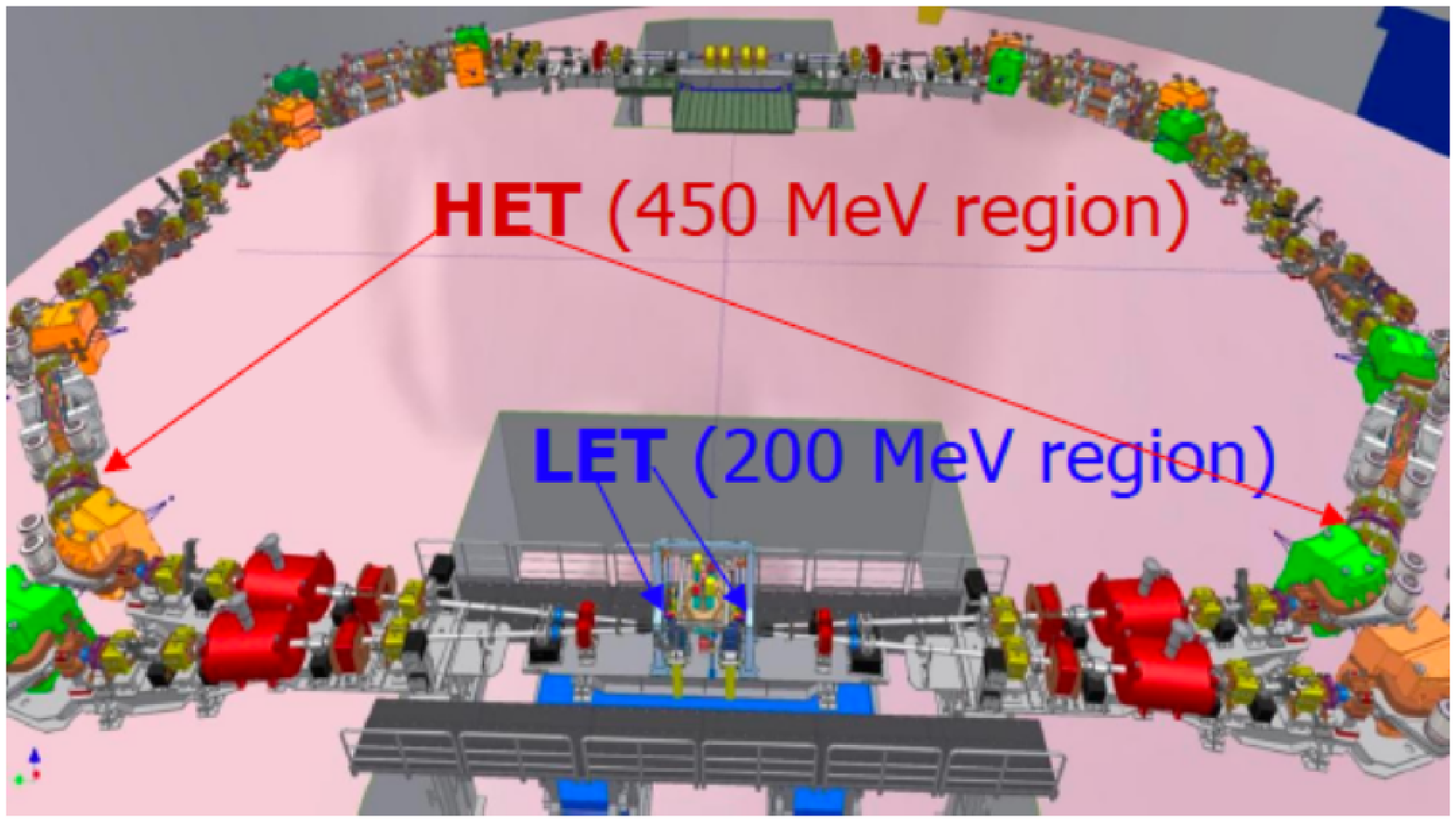}
\caption{ (Left) Cross section of the KLOE detecor.
  (Right) Scheme of the DA${\Phi}$NE collider with positions
          of Low Energy Taggers (LET) and High Energy Taggers (HET)
          indicated by arrows. The average energy of electrons and positrons covered is also shown.
  \label{kloedetector}}
\end{figure}

The KLOE-2 experimental setup~\cite{kloe2amelino,kloe2_fabio}
is a successor of KLOE~\cite{kloe,EMCkloe,DCkloe},
which is at present being upgraded by
new components
in order to improve its tracking and clustering
capabilities as well as in order
to tag $\gamma\gamma$ fusion processes.
The detector, shown schematically in the left panel of Fig.~\ref{kloedetector}
consists of a $\sim$~3.5~m long cylindrical drift chamber with a diameter of about 4~m
surrounded by the sampling electromagnetic calorimeter~\cite{kloe,EMCkloe,DCkloe}.
Both these detectors
are immersed in the axial magnetic field
provided by the superconducting solenoid.  
The detector surrounds the crossing of the positron and electron beams
circulating in the rings of the DA$\Phi$NE collider. 
Each out of 120 bunches of positrons and electrons collides with its counterpart 
once per turn in the center of the KLOE-2 detector~\cite{flavorofKLOE}.
Last year the new  electron-positron interaction region, based on
large Piwinski angle, small beam sizes at the crossing point,
and {\em Crabbed Waist} compensation of the beam-beam interaction
has been successfuly commissioned~\cite{zubov}.
This new solution allowed to increase  the collider
luminosity by a factor of three with respect to the performance reached before
the upgrade,  and DA$\Phi$NE will deliver up to 15 pb$^{-1}$ per day giving
possibility to achieve about 20~fb$^{-1}$ within the next 3-4 years of data taking by means of the
KLOE-2 detector.
In addition exclusive measurements of the $\gamma\gamma$ reactions will be enabled by the low
and high energy taggers~\cite{LET} allowing for registration of electrons and positrons
originating from  $e^{+}e^{-}\to e^{+}e^{-}\gamma^*\gamma^*\to e^{+}e^{-}X$ reaction.
Right panel of Fig.~\ref{kloedetector}
shows the scheme of the DA${\Phi}$NE rings with position of taggers indicated by arrows.
First commissioning runs with KLOE-2 started in spring this year and,
after collection of statistics corresponding to the integrated luminosity of about 5~fb$^{-1}$,
the next phase of installation of new detectors including inner tracker and internal calorimeters
shall commence
by the end of the year 2011.

\section{WASA at COSY}
WASA-at-COSY~\cite{wasaatcosyproposal}  is a successor of WASA/CELSIUS experiment 
which was operated 
until  2005  at   the  CELSIUS  light  ion   storage  ring in Uppsala
\cite{Bargholtz:2008ze}.
The detector was optimized for studies of $\pi^0$ and $\eta$
decays  involving   photons  and   electrons~\cite{Kupsc:2009zz1}.
 In the year 2006 it was installed at the beam of the cooler 
synchrotron COSY in J{\"u}lich~\cite{cosy,cooling}. 
Cooler synchrotron COSY is equipped with electron and stochastic cooling providing 
low emittance polarized and unpolarized proton and deuteron beams with momentum 
of up to 3.7 GeV/c. 
\begin{figure}[h] 
       {\centerline{\includegraphics[width=11cm]{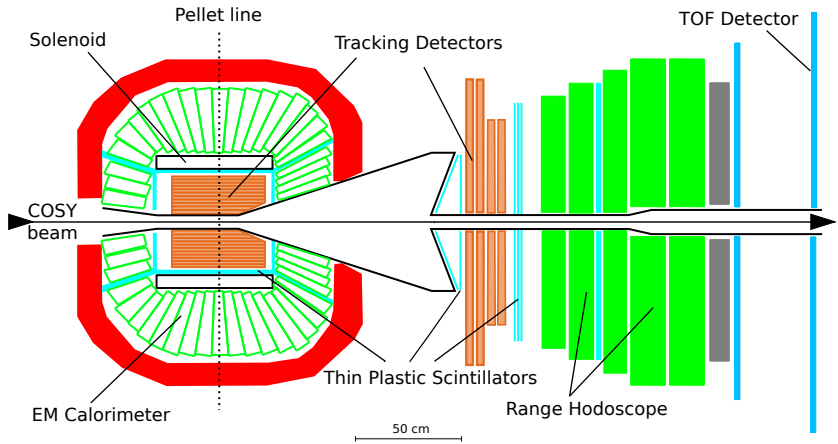}}}
\caption{ 
   Scheme of the WASA detector setup installed at COSY.
  \label{wasadetector}}
\end{figure}
Therefore the transfer of the WASA detector 
from CELSIUS to COSY opened possibilities
to study the production and decays of mesons heavier than the $\eta$ 
with masses up to the mass of the $\phi$ meson, and permitted to extend the studies
to the spin degrees of freedom. It is also important to stress that COSY allows for the
continuous variation of the beam  momentum within one cycle which is crucial for reduction
of uncertainties 
in studies 
of excitation functions of the meson production processes.
The WASA detector system (shown schematically in Fig.~\ref{wasadetector}) consists of 
the Forward Detector used for tagging of meson production, the Central Detector used for the registration 
of the decay products, and the pellet target  system.
The production of the $\eta$ mesons
occurs in the middle of the Central Detector in the intersection of COSY beam with the vertical 
beam of pellets. The interaction region is 
surrounded by the multi-layer cylindrical drift chamber
immersed in the axial magnetic field produced by the superconducting solenoid. 
The outermost sensitive
part of the Central Detector is the electromagnetic calorimeter covering 96 percent of the whole solid angle.
Particles registered in the Forward Detector are identified based on the energy loss 
in the layers of the scintillator detectors and their direction of flight is reconstructed based on signals measured
in the multi-layer drift chambers. 

The production of the $\eta$ meson is conducted only a few tens 
of MeV above the kinematical threshold. 
The relatively small excess energy allows for the efficient separation
of
decay products emitted into a large solid angle
from the forward boosted protons and helium ions.

\section{Tagging of $\eta$ mesons}
\begin{figure}[h]
       \vspace{-0.3cm}
       \includegraphics[width=7.1cm,height=6.0cm]{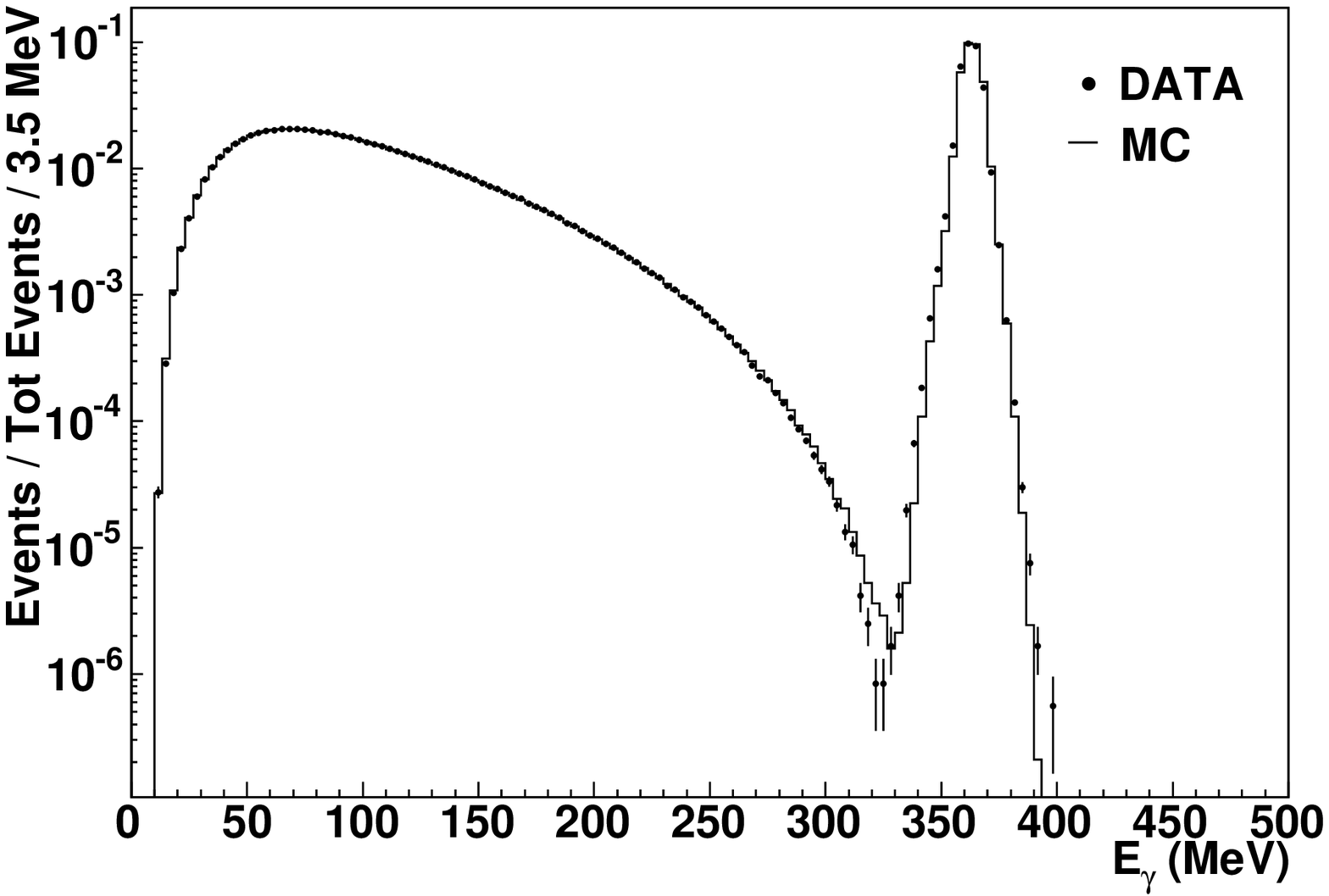}\hspace{-0.6cm}
       \includegraphics[width=6.0cm,height=6.7cm,angle=90.]{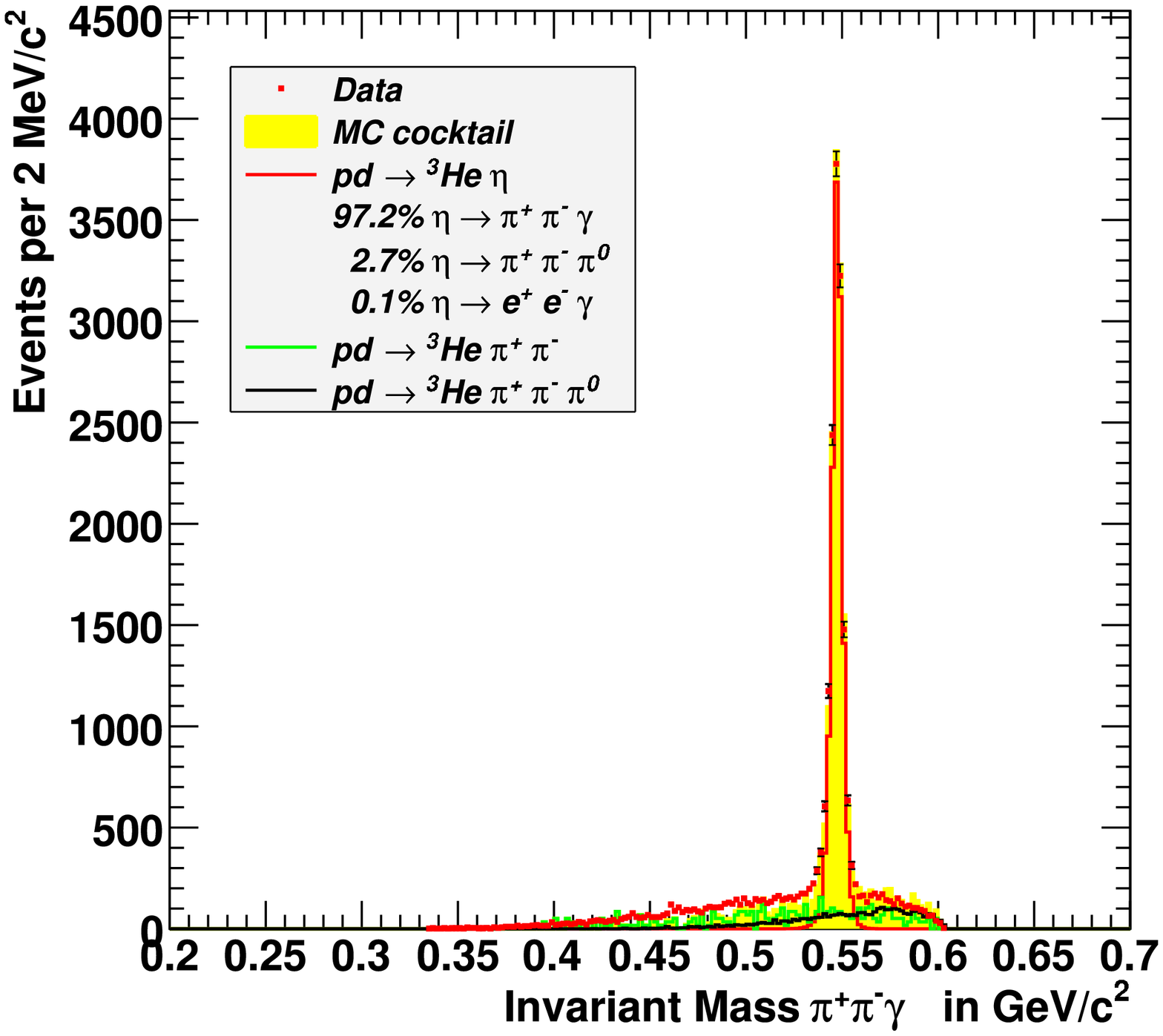}
\caption{(left) KLOE: Energy spectrum for $\gamma$ quanta  from  events 
$\phi \to \eta \gamma \to \pi^{0} \pi^{0} \pi^{0} \gamma$, showing the
363-MeV tagging photon well separated from those from $\pi^0$ decay~\cite{caterina}.
(Right) WASA-at-COSY: Missing mass spectrum for the 
$pd\to ^3\!\!\!He \, X \to ^3\!\!\!He \, \pi^+ \pi^- \gamma$~\cite{christophPhD}, 
showing a clear maximum at mass value equal to the mass of the $\eta$ meson.
  \label{tagging}}
\end{figure}

At the DA$\Phi$NE collider the $\eta$  mesons are created in the center of the KLOE-2 detector
via radiative decays of 
$\phi$ meson produced in the electron-positron collisions ($e^+e^- \to \phi \to \eta \gamma$)
and via the fusion of virtual gamma quanta exchanged in the $e^+ e^-$ interaction 
($e^+ e^- \to e^+  e^- \gamma^* \gamma^* \to e^+ e^- \eta$).  The production of $\eta$ meson
is tagged respectively via the registration of mono-energetic gamma quantum in the calorimeter  
or via registration of electrons and positrons in low and high energy taggers.
Left panel of Fig.~\ref{tagging} presents the capability of the KLOE detector for the
clear identification of the $\eta$ meson via the detection of the
monoenergetic $\gamma$ quantum
from the $\phi \to \eta \gamma$  decay.

At the cooler synchrotron COSY the $\eta$ mesons are produced in the middle of the central part
of the WASA detector in collisions 
of circulating proton or deuteron beam with the stream of frozen hydrogen or deuterium pellets.
In particular $pp\to pp\eta$ and $pd\to ^3\!\!He \, \eta$ reactions are used and  
the production of the $\eta$ meson is tagged 
by the registration of protons or helium ions in the forward part of the WASA detector.
An example of the missing mass spectrum of the $pd\to ^3\!\!\!He \, X $  reaction
showing the capability of the WASA-at-COSY detector setup
is presented in the right panel of Fig~\ref{tagging}. 

\section{Discrete symmetries and QCD anomalies}
\begin{figure}[h]
      \vspace{-0.4cm}
      {\centerline{  \includegraphics[height=0.17\textwidth]{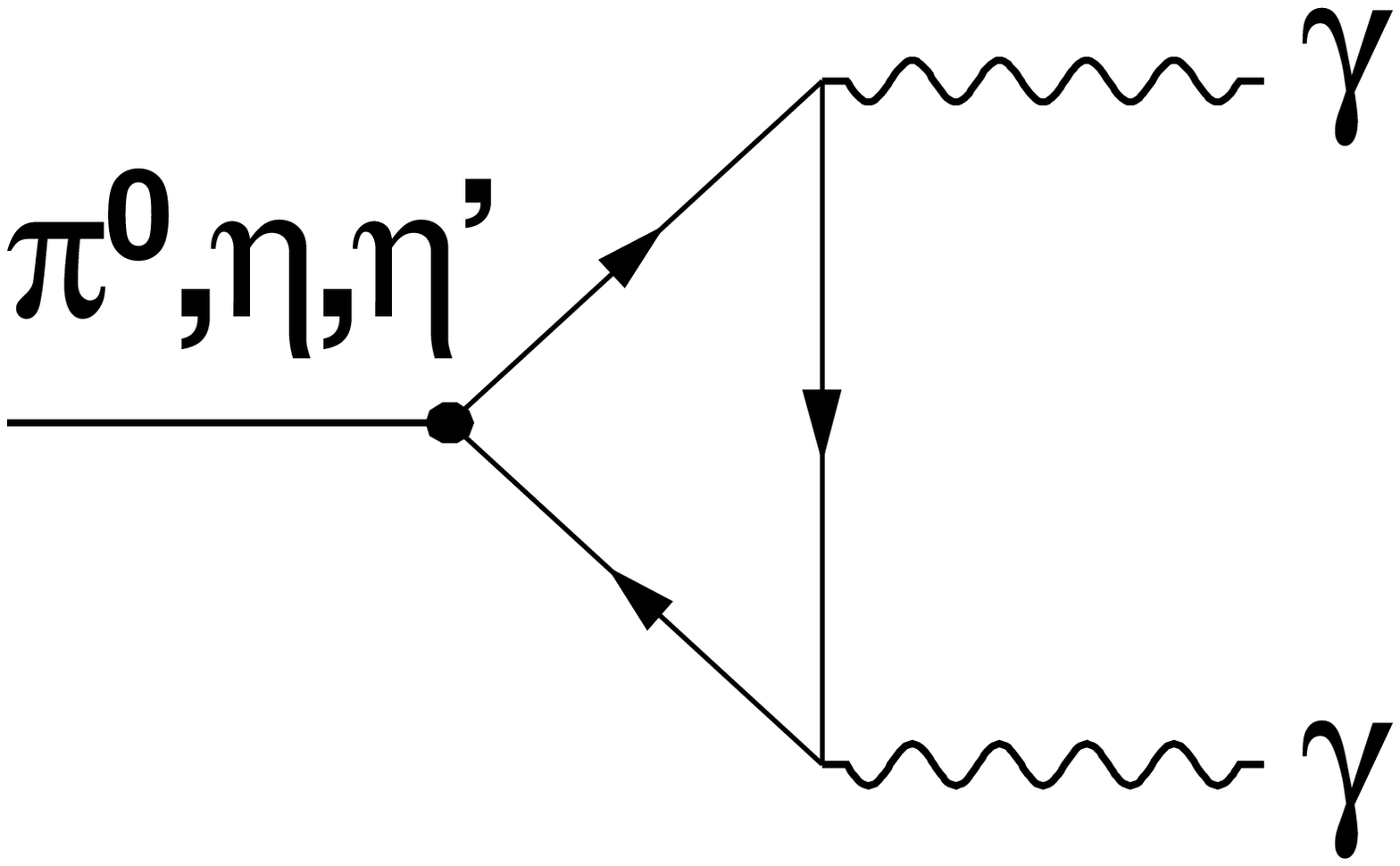} 
        \includegraphics[height=0.17\textwidth]{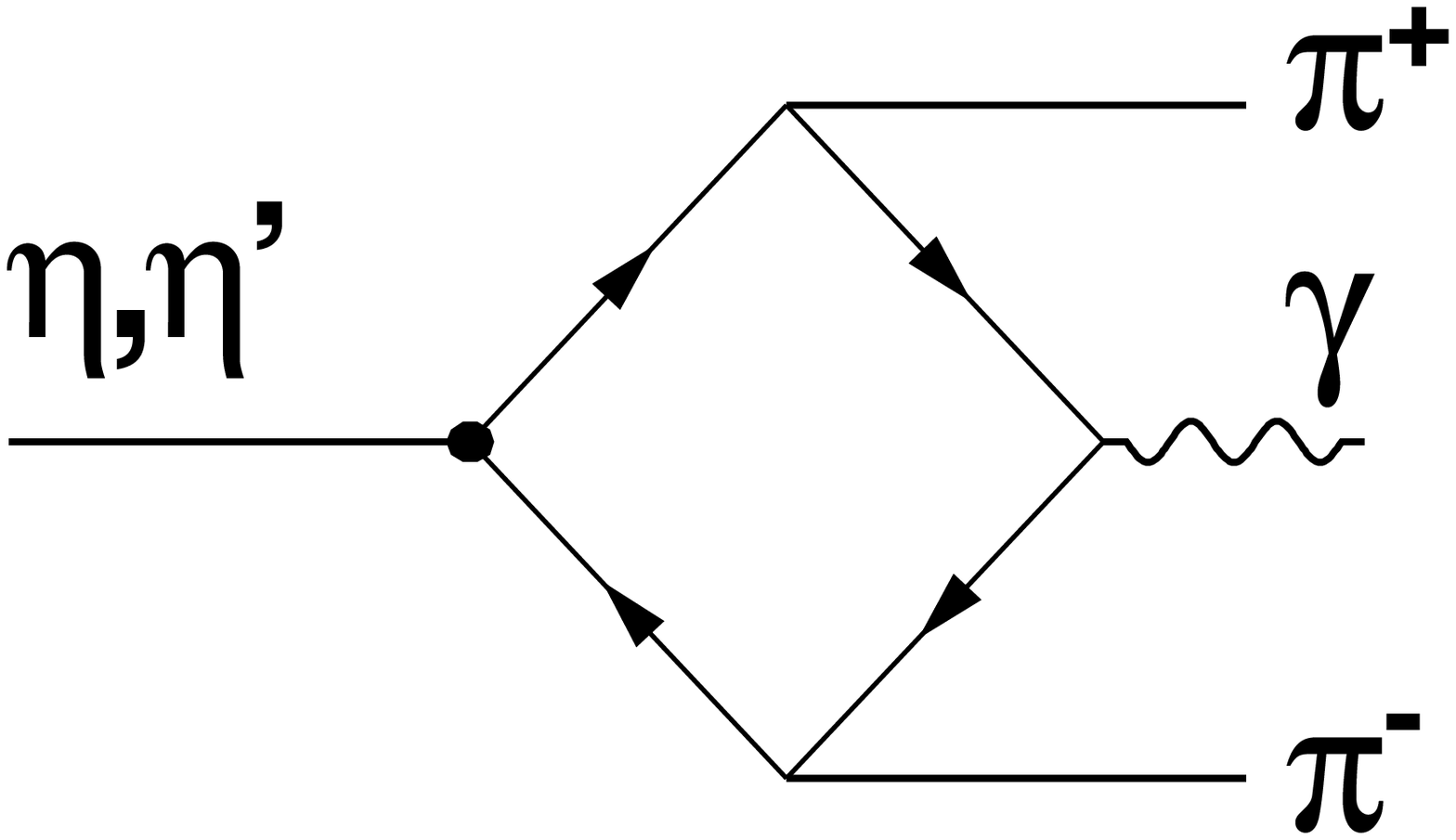} }}
        \vspace{-0.5cm}
       \caption{Diagrams for the triangle and box anomalies.}
     \label{box_triangle}
    \end{figure}

\begin{figure}[h]
\vspace{-0.3cm}
  {\centerline{ \includegraphics[height=0.16\textwidth]{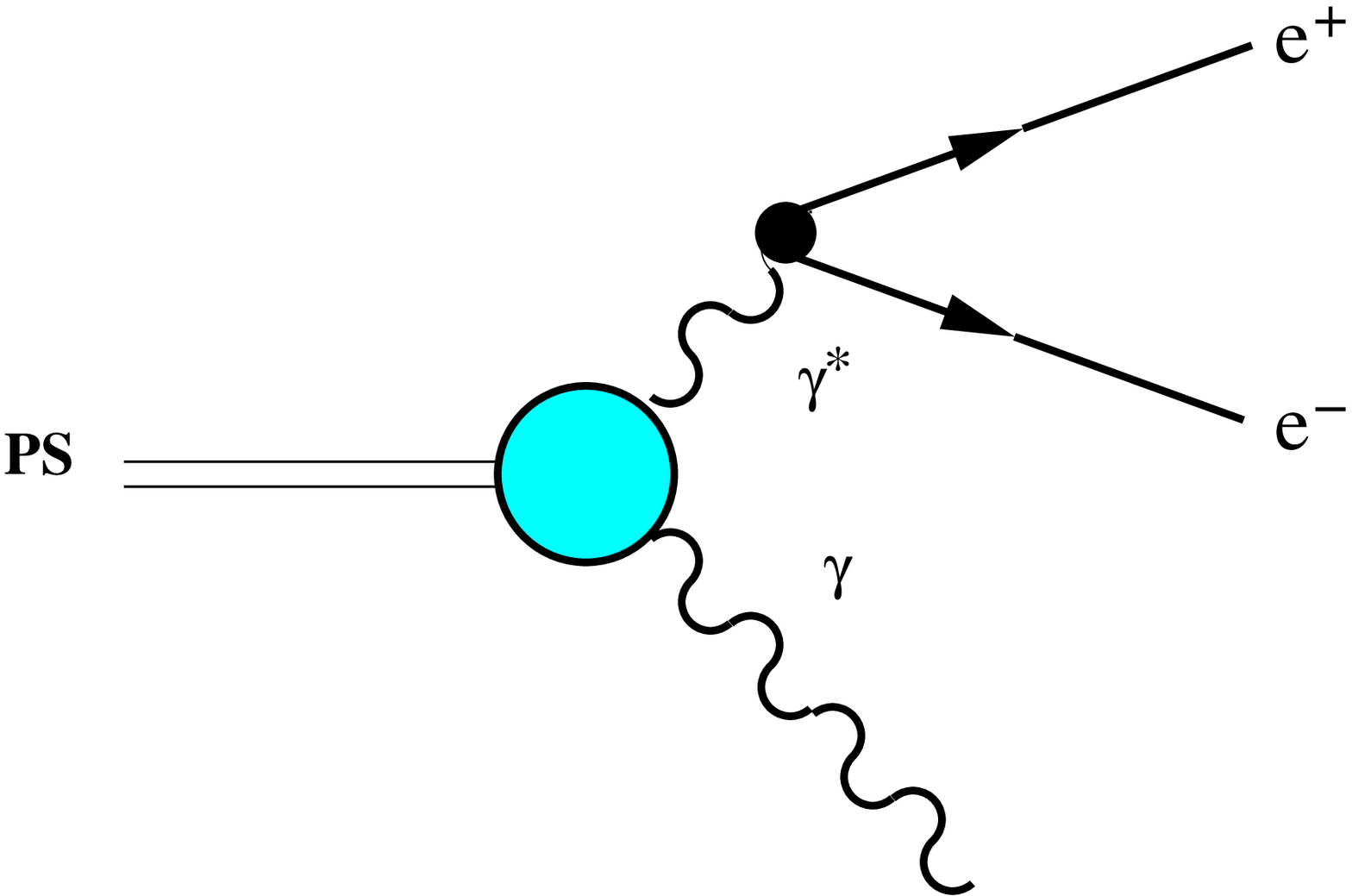} \hspace{2cm}
    \includegraphics[height=0.16\textwidth]{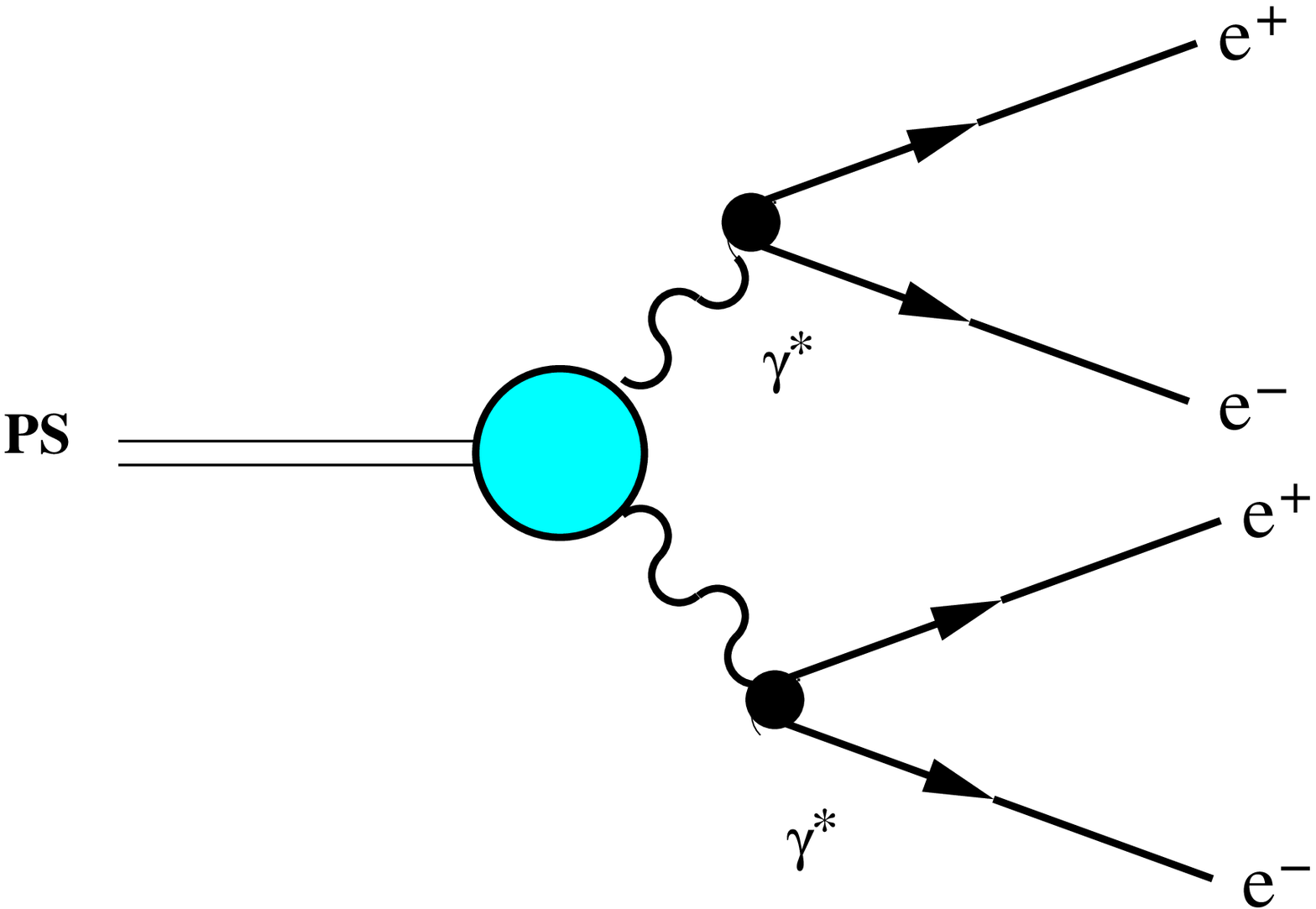} }}
\vspace{-0.3cm}
  \caption{
  Diagram for Dalitz (left) and double Dalitz
    decays (right) of pseudoscalar mesons.}
  \label{dalitz}
\end{figure}

    Both $\eta$ and $\eta^{\prime}$ mesons are eigenstates of operators of parity (P), charge conjugation (C) 
    and combined  CP parity (with eigenvalue of P~=~-1, C~=~+1 and CP~=~-1). 
    Therefore, studies of their decays constitute a valuable source of information  
    regarding the degree of conservation of these symmetries in strong and electromagnetic interactions. 
    In this context particularly interesting is the $\eta$ meson
    since all its strong and electromagnetic decays are forbidden in the first order~\cite{nefkens}. 
    The most energetically favourable strong decay of $\eta$ into  $2\pi$ is forbidden 
    due to P and CP invariance. 
    Its decay into $3\pi$ is suppressed by G-parity and isospin invariance~\cite{marcinmgr} 
    and it occurs due to the  difference between the mass of u and d quarks  thus enabling the 
    study of these masses since electromagnetic effects are expected to be small~\cite{miller,sutherlandPL}.
    Further on, strong decay into $4\pi$ is suppressed due to the small available phase space 
    and again due to P and CP invariance.
    The first order electromagnetic decays as $\eta \to \pi^0 \gamma$  or $\eta \to 2\pi^0 \gamma$  break 
    charge conjugation invariance and  $\eta \to \pi^+ \pi^- \gamma$ 
    is also suppressed because charge conjugation 
    conservation requires odd (and hence nonzero) angular momentum in the $\pi^+ \pi^-$ system. 
    Moreover, this radiative decay at a massless quark limit 
    is driven by the QCD box anomaly~(Fig~\ref{box_triangle}).
    In addition, in the massless quarks limit also the second 
    order electromagnetic decay $\eta \to \gamma \gamma$ 
    is forbidden~\cite{sutherland,veltman,nefkens}, 
    and it accurs only 
    due to the QCD triangle anomaly~\cite{adler,bell}~(Fig.~\ref{box_triangle}).
    The above mentioned features of the $\eta$ meson makes it especially 
    suitable for the test of discrete 
    symmetries in strong and electromagnetic interactions and for investigations of QCD anomalies. 
    The QCD triangle 
    and box anomaly 
    involved correspondingly in the $\eta\to \gamma\gamma$ and $\eta\to \pi^+\pi^-\gamma$ 
    decays may also be studied 
    via leptonic and semi-leptonic decays of 
    meson  
    in which the virtual $\gamma$ quantum
    converts internally into $e^+e^-$ pair e.g. via Dalitz or double Dalitz decays: 
    $\eta\to \gamma^* \gamma \to e^+e^-\gamma$,
    $\eta\to \gamma^* \gamma^* \to e^+e^- e^+e^-$ as shown in Fig.~\ref{dalitz} or  
    via semi-leptonic decay as eg.  $\eta\to \pi^+\pi^-\gamma^* \to \pi^+\pi^- e^+e^-$.  
     The $\eta\to e^+e^-\pi^+\pi^-$ process  
     is interesting also because it
     enables to study
     the flavour-conserving CP violation not predicted in the framework of the Standard Model,
     and not constrained by the experimental limits on $\eta\to \pi\pi$ decays or on the electric
     dipole moment of the neutron.
     The violation of the CP symmetry in the $\eta\to e^+e^-\pi^+\pi^-$ decay
     would manifest itself as an angular asymmetry between pions and electrons
     decay planes. A possible mechanism leading to 
     such asymmetry could be an interference between the electric 
     and magnetic transition leading to the  linear polarisation of the $\gamma$ quantum 
     from the $\eta \to \pi^+\pi^-\gamma^* \to \pi^+\pi^- e^+ e^-$ process~\cite{gao,geng}.  
     In the case of the  $K_L$  
     meson such asymmetry was observed by  the
     KTeV
    collaboration~\cite{AlaviHarati:1999ff}.   
\begin{figure}[h]
\vspace{-0.5cm}
  {\centerline{
   \includegraphics[height=0.2\textwidth]{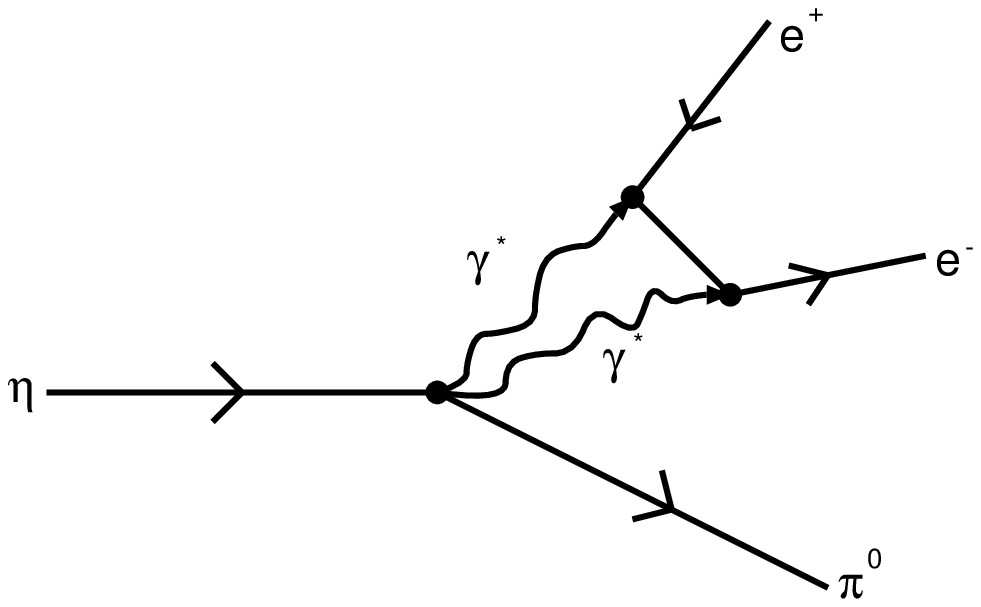} \hspace{2cm}
    \includegraphics[height=0.2\textwidth]{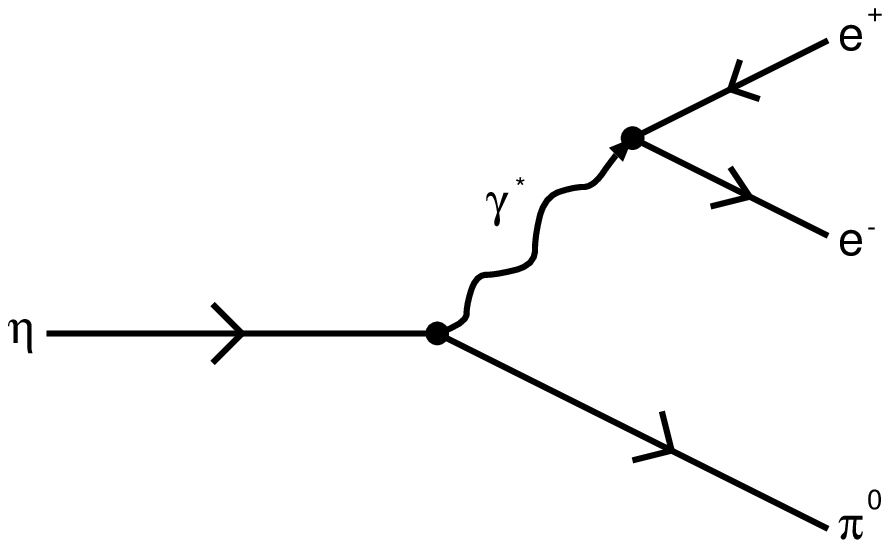}}}
\vspace{-0.5cm}
  \caption{
(Left) Second order electromagnetic C-conserving transition: $\eta\to\pi^0\gamma^*\gamma^*\to\pi^0 e^+e^-$.
(Right) Diagram for the C invariance violating first 
order electromagnetic process: $\eta\to\pi^0\gamma^*\to\pi^0 e^+e^-$.
  \label{pi0ee}
}
\end{figure}

      As regards the C symmetry conservation so far it was not so extensively investigated 
      in electromagnetic and strong interactions. It may be tested in $\eta$ decays 
      into odd numbers of gamma quanta
      such as e.g. $\eta \to 3\gamma$, $\eta \to \pi^0\gamma$, $\eta \to 3\pi^0 \gamma$ etc.
      Also the conversion decays as $\eta\to\pi^0 e^+e^-$ may be used for the test of charge conjugation invariance.
      As shown in Fig.~\ref{pi0ee}, in the framework of the Standard Model, 
      this process may proceed via C-conserving exchange of two virtual $\gamma$ quanta 
      ($\eta \to \pi^0 \gamma^* \gamma^* \to \pi^0 e^+e^-$) with the branching ratio 
      of about $10^{-8}$, but in principle it may also be realized with only one $\gamma$ quantum 
      in the intermediate state (Fig~\ref{pi0ee}), 
      thus breaking C invariance and increasing the branching ratio.     
      At present the empirical upper limit determined for the 
      BR($\eta\to\pi^0 e^+e^-$) amounts to 4.5~x$10^{-5}$.
      Present upper limits for other C-violating decays ( e.g. 
       BR($\eta\to\gamma\gamma\gamma) <  1.6~10^{-5}$~\cite{kloeAloisio})
      are at the same level. 
      One may also test the C invariance in hadronic decays  as e.g. $\eta \to \pi^0 \pi^+ \pi^-$ where 
      it can manifest itself as an asymmetry in the energy distributions for $\pi^+$ and $\pi^-$ mesons 
      in the rest frame of the $\eta$ meson. The studies of the asymmetries in the population 
      of the Dalitz plot for this decay permits also to study the isospin  symmetry breaking 
      and test the Chiral Perturbation 
      Theory~\cite{wasaatcosyproposal,Adolph:2008vn,Kupsc:2009zz,Kupsc:2009zz1}.

Both KLOE-2 and WASA-at-COSY  aim at a
significant improvement of the sensitivity of the tests of the
discrete symmetries
in the decays of $\eta$ and $\eta^{\prime}$
mesons beyond the presently achieved limits. In some cases 
like e.g. the tests of $P$, $C$, or $CP$ symmetries,
with an expected number of about $10^9$ $\eta$ tagged, 
an improvement by more than one order of magnitude is expected.

\section{Spatial distributions of $\eta$ and $\eta^{\prime}$}
Due to the short life time of the flavour neutral mesons their structure as well as interaction cannot be studied
by means of classical scattering experiments. In addition, due to the positive charge conjugation 
of pseudoscalar mesons their structure cannot be  studied based on processes with the exchange of one photon.
Therefore in order to investigate the spatial distribution of the meson charge one has to study its decays 
into two photons out of which  at least one is virtual as it is e.g. in Dalitz or double Dalitz decays. 
The internal conversion of the virtual photon leads to the creation of the  $l^+ l^-$ pair 
with a square of the invariant mass equal to the square of the four-momentum 
vector of the virtual photon ($q^2$).
Deviations of the $l^+ l^-$ invariant mass distributions 
from the predictions based on the assumption of point like meson
(which may be conducted in the framework of the Quantum Electrodynamic) deliver information about the meson structure
and the dynamics of the process. These deviations are characterized by means of 
the transition form factors which in general depend on 
the square of the four-momentum of the involved photons. 
\begin{figure}[h]
   {\centerline{\includegraphics[height=0.15\textwidth]{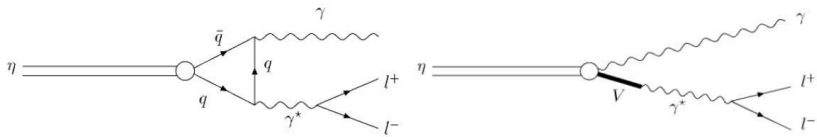}}}
   \vspace{-0.4cm}
  \caption{
 Diagram representing the decay of the $\eta$ meson into $\gamma$ quantum and lepton anti-lepton pair:
(left) in the framework of the quark model and (right) in the framework of the vector meson dominance model (VDM)~\cite{Stepaniak}.
  \label{VDM}
}
\end{figure}
Distributions of the transition form factors as a function of $q^2$ 
allow 
to investigate the contribution to the process from e.g. triangle QCD anomaly 
or vector dominance mechanism (see Fig.~\ref{VDM}). The studies of the 
conversion decays gives information about the time-like region of the form-factor 
with positive $q^2$ 
equal to the square of the invariant mass of the $l^+ l^-$ pair.
The information about the space-like region with the negative values of $q^2$ is accessible 
via cross section of mesons production in  $\gamma^* \gamma^*$ fusion realized in e.g. 
$e^+ e^- \to e^+ e^- \gamma^*\gamma^* \to e^+ e^- \eta$ reaction. 
It is important to stress that the knowledge
of the transition form factors for pseudoscalar mesons 
is of crucial importance for the studies 
of the anomalous magnetic moment of the muon 
$a_{\mu} = (g_{\mu}-2)/2$  
which constitutes one of the most precise 
test of the Standard Model. $a_{\mu}$ was measured with the 
precision of 
0.5 ppm~\cite{Bennett:2006fi}
and 
FNAL experiment~\cite{Carey:2009zz}
plans to improve this accuaracy to 0.14 ppm in the near future. 
\begin{figure}[h]
   \vspace{-0.4cm}
  {\centerline{ \includegraphics[height=0.15\textwidth]{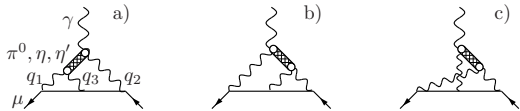}}}
   \vspace{-0.2cm}
  \caption{
Pseudoscalar exchange contribution to the light by light
  scattering.
  \label{lbl}
}
\end{figure}
The predictions of the value of $a_{\mu}$ based on the SM are however limited by the accuracy 
of the determination of the hadronic contributions.  The accuracy 
of pseudoscalar transition form factors 
dominates the precision in determination of hadronic light-by-light contributions via terms shown in Fig~\ref{lbl},
and therefore the precise studies of transition form factors 
is of importance for the SM predictions of the
anomalous magnetic moment of the muon.

\section{Search for the U  boson}
The measurement of radiative decays of mesons (in particular of the $\eta$ meson) 
is interesting also in a context of the search for the explanation 
of the recent astrophysical observations of  excess of $e^+ e^-$ annihilation $\gamma$ quanta
from the galactic center observed by
the INTEGRAL satellite~\cite{Jean:2003ci}, the excess in the
cosmic ray positrons reported by PAMELA~\cite{Adriani:2008zr}, the total
electron and positron flux measured by ATIC~\cite{Chang:2008zzr}, Fermi~\cite{Abdo:2009zk},
and HESS \cite{Aharonian:2009ah},
and the annual modulation of the DAMA/LIBRA signal~\cite{Bernabei:2008yi}.
In one of the considered scenario which may be reconciled with the observations~\cite{kloe2amelino}, 
the origin of this kind of enhanced stream of radiation, 
may be explained assuming
that positrons are created in an annihilation of the dark matter particles into $e^+ e^-$ pairs,
and that this process is mediated by the  
U boson with mass in the GeV 
scale~\cite{Boehm,Fayet,Boehm2,kloe2amelino,arkani}. 
The existence of such U boson could manifests itself 
as a maximum in the invariant mass distribution of $e^+ e^-$ pairs originating 
from the radiative decays 
as e.g. $\eta \to \gamma U \to \gamma e^+ e^-$~\cite{reece}.
 In addition, there are several
possible signatures of U boson which may be searched for at $e^+ e^-$ collider DA$\Phi$NE,
 such as  $e^+e^-\to e^+e^-+\gamma$,
 $e^+e^-\to \mu^+\mu^-+\gamma$, $e^+e^-\to E_{missing}+\gamma$,
 $e^+e^-\to E_{missing} + e^+e^-$, or events with 4 or 6 leptons in
the final state~\cite{kloe2amelino}. 
Existence of such boson would also influence the branching ratio for 
rare processes such as e.g. $\pi^0\to e^+ e^-$
and $\eta \to e^+ e^-$. The latter process was never 
observed till now.   The present empirical upper limit set
by the WASA/CELSIUS experiment~\cite{Berlowski} 
is still by four orders of magnitude larger than SM predictions (~$10^{-9}$).
Small probability for this decay make it especially sensitive 
to any hypothetical interactions from beyond the Standard Model. 
For detailed overview of phenomenology and possibilities of the search 
for the dark matter particles in the low energy facilities at the GeV scale 
the interested reader is referred to article~\cite{kloe2amelino}.

\section{Search for the eta-mesic nuclei}
The negatively charged pions and kaons can be trapped
in the Coulomb potential of atomic nucleus  forming so called pionic (kaonic) atoms.
Observations of such atoms allows for studies of strong interaction of pions and kaons
with atomic nuclei on the basis of shifts and widths of the energy levels~\cite{strauch,zmeskal}.
It is also conceivable that a neutral meson could be bound to a nucleus.
In this case the binding is exclusively due to the strong interaction
and hence such object can be called a {\it mesic nucleus}.
Here the most promising candidate
is the $\eta$-mesic nucleus since the $\eta$N interaction
is strongly attractive.
We may thus  picture~\cite{sokol0106005} the formation and decay of the
$\eta$-mesic nucleus
as the $\eta$ meson absorption by one of the nucleons
leading to the creation of the $N^*(1535)$ and
then its propagation through the nucleus until it decays into
the pion-nucleon pair
which escapes from the nucleus.
Predicted values of the width of such states range
from $\sim$7 to $\sim$60~MeV~\cite{osetPLB550,liu3,haideracta,Tsushima:1998qw}.
The search of the $\eta$~-~mesic nucleus was conducted
in many inclusive experiments
via reactions induced by pions~\cite{bnl,lampf}, 
protons~\cite{jinr,gem,Machner:2010hs,Moskal:2010ee,Khoukaz:2010dw}, 
and photons~\cite{lpi,mami,Krusche:2010du}.
Many  promising indications
of the existence of such an object
were reported,
but so far none was independently confirmed.
Experimental investigations with high statistical sensitivity
and the detection of the N$^*(1535)$ decay products are being continued at
the COSY~\cite{Moskal:2010ee},
JINR~\cite{jinr}, J-PARC~\cite{Fujioka:2010dv}, and
MAMI~\cite{Krusche:2010du} laboratories.
For comprehensive description of the interpretation of present results 
the interested reader is referred to the recent 
theoretical publications from the international symposium 
on mesic-nuclei~\cite{Haider:2010dt,Bass:2010kr,Wilkin:2010am,Wycech:2010dm,Hirenzaki:2010dn}.
Here we only report on the searches of the $\eta$~-~mesic helium
carried out by means of the WASA detector at COSY~\cite{Moskal:2010ee}.
We  conduct a search
via an exclusive measurement
of the excitation functions for the $dd \rightarrow {^3\mbox{He}}\, p\, \pi^-$ 
and 
 $dd\rightarrow ^{3}\hspace{-0.03cm}\mbox{He} n \pi{}^{0} \rightarrow$ $^{3}\hspace{-0.03cm}\mbox{He} n \gamma \gamma$
reaction
varying  continuously
the beam momentum
around the
threshold for the $d\,d \to ^4\!{\mathrm{He}}\,\eta$ reaction.
Ramping of the beam momentum and
taking advantage of the large
acceptance of the WASA detector
allows to minimize systematical uncertainties making
the WASA-at-COSY a unique facility~\cite{Moskal:2010ee}
for such kind of exclusive experiments.
The  ${^4\mbox{He}}-\eta$
bound state should manifest itself as a resonant like structure
below the threshold for the $dd \to {^4\mbox{He}}\,\eta$ reaction.

In the first experiment conducted in June 2008, no structure was found which would indicate 
the existence of the eta-mesic helium,
and an upper limit for the eta-mesic helium production via
the $dd \to (\eta\,{^4\mbox{He}})_{bound} \to {^3\mbox{He}}\, p\, \pi^-$ reaction 
was determined to be  about
20~nb on a one sigma level.
In November 2010 the statistics was increased by 
a factor of about 40 and the data are now under evaluation.

\section{Gluonium content}
     In the quark model the $\eta$ and $\eta^{\prime}$ mesons are regarded as the mixture
of the singlet 
and octet 
states of the SU(3)-flavour pseudoscalar nonet.
A small pseudoscalar mixing angle 
implies
that the percentage amount of various quark flavours in both mesons is almost the same~\cite{hab}.
Nevertheless, their physical properties are unexpectedly different.
The $\eta^{\prime}$(958) meson mass
is almost two times larger than the mass of the $\eta$(547) meson.
The branching ratios for the decays of $B$ and $D_s$ mesons
into the $\eta^{\prime}$ meson exceed significantly those into the $\eta$ meson
and the standard model predictions, 
especially in processes requiring involvement of gluons~\cite{ball-plb365,fritzsch-plb415}.
There exist excited states of nucleons
which decay via emission of the $\eta$ meson, 
yet none of the observed baryon resonances decays via  the
emission of the $\eta^{\prime}$ meson~\cite{PDG}. 
Their production cross sections in the collisions of nucleons
differs by more than order of magnitude~\cite{hab,Moskal:2002jm}
and also their hadronic interaction with nucleons differs significantly~\cite{Moskal:2000pu}. 
Due to the small mixing angle the $\eta^{\prime}$ meson remains predominantly 
the SU(3)-flavor singlet and therefore it may include
pure gluon states  to a much larger extent than the $\eta$  
and all other pseudoscalar and vector mesons~\cite{stevenActaSup}.

Recently the gluonium fraction of  $(12 \pm 4)$\% in the $\eta^{\prime}$ meson 
has been extracted fitting the widths of the magnetic
dipole transition $V \to P \gamma$, where $V$ are the vector mesons
$\rho, \omega,\phi$ and $P$ the pseudoscalar mesons $\pi^0, \eta, \eta'$,
together with the $\pi^0 \to \gamma \gamma$ and $\eta' \to \gamma \gamma$
partial widths, and using 
the KLOE measurement of
$R_{\phi} = BR(\phi \to \eta' \gamma)/BR(\phi \to \eta \gamma)$~\cite{Ambrosino:2006gk}.
The accuracy of the fit depends among others also on the uncertainty of the natural width  of the $\eta^{\prime}$ meson.
The total width of the $\eta^{\prime}$ meson extracted by PDG~\cite{PDG}
is strongly correlated with the value of
the partial width $\Gamma(\eta'\to\gamma\gamma)$~\cite{PDG},
which causes serious difficulties when the total and the partial width have to be used at the same time,
like e.g. in studies of the gluonium content of the $\eta^{\prime}$ 
meson~\cite{Biagio}.
In this context it should be mentioned that the
inaccuracy  of $\Gamma_{\eta'}$ limits not only the extraction of the $\eta^{\prime}$ gluonic content but it limits also
investigations of many other interesting physics issues,
as for example the quark mass difference $m_{d}-m_{u}$~\cite{Borasoy2,marcinmgr,Kupsc},
isospin breaking in Quantum Chromo-Dynamics (QCD)~\cite{Borasoy,Borasoy2}, or
the box anomaly of QCD~\cite{Nissler}.
This is because the branching ratios of the $\eta^{\prime}$ meson decay channels are typically
known with a relative precision of more than
an order of magnitude better than the present
accuracy with which $\Gamma_{\eta'}$ 
is extracted~\cite{PDG}~\footnote{
Recently the COSY-11 collaboration extracted 
the total width of the $\Gamma_{\eta'}$ directly 
from the mass spectra
and obtained: 
$\Gamma_{\eta'}=0.226\pm0.017(\textrm{stat.})\pm0.014(\textrm{syst.})$~MeV/c$^2$~\cite{eryk}.
The result does not depend on knowing any of the branching ratios or partial decay widths.}.

With the KLOE-2 data-taking above the $\phi$ peak, e.g., 
at  $\sqrt{s} \sim 1.2$ GeV,  
it is possible to measure the $\eta'$
decay width $\Gamma(\eta' \to \gamma \gamma)$ through 
$\sigma(e^+ e^- \to e^+ e^- (\gamma^* \gamma^*) \to e^+ e^- \eta')$.
The measurement to 1\% level of both the cross section and the
$BR(\eta' \to \gamma \gamma)$ 
would bring the fractional error on the 
$\eta'$ total width,
$\Gamma_{\eta'} = \Gamma(\eta' \to \gamma \gamma) /
BR(\eta' \to \gamma \gamma)$,
to $ \sim$1.4\%~\cite{kloe2amelino}, 
and can reduce the uncertainty of the gluonium content determination by about 
a factor of three.

The gluonic admixture of $\eta^{\prime}$  
influences also the $\eta^{\prime}$-nucleon interaction and production processes
via the U(1) anomaly~\cite{bass286}.
The range of the glue induced $\eta^{\prime}$-nucleon interaction,
if determined by the two-gluon effective potential, would be in the order of 0.3~fm~\cite{baru445}.
This range is large enough
to be important in the threshold production of the $\eta^{\prime}$ meson e.g. via the $pp\to pp\eta^{\prime}$ reaction
which occurs at distances of the colliding nucleons in the order of 0.2~fm. At such small distances the quark-gluon
degrees of freedom may play a significant role in the production dynamics of the $\eta$ and $\eta^{\prime}$ mesons.
Therefore, additionally to the mechanisms associated with meson
exchanges it is possible that the $\eta^{\prime}$ meson is created from excited glue
in the interaction region of the colliding nucleons~\cite{bass286,bass348},
which couple to the $\eta^{\prime}$ meson directly via its gluonic
component or through its SU(3)-flavour-singlet admixture. The production through the
colour-singlet object as suggested in reference~\cite{bass286} is isospin independent
and should lead to the same production yield of the
$\eta^{\prime}$ meson in the $pn\to pn\ gluons \to pn\eta^{\prime}$
and                          $pp\to pp\ gluons \to pp\eta^{\prime}$ reactions
after correcting for the final and initial state interaction between the nucleons.
Such studies were conducted in the case of the $\eta$ meson~\cite{calen-prc58,moskal-prc79}.
The ratio
$R_{\eta} 
 = \sigma (pn \rightarrow pn \eta ) / \sigma (pp \rightarrow pp \eta )$
has been measured for quasifree $\eta$
production from a deuteron target up to 109 MeV 
above threshold~\cite{calen-prc58,moskal-prc79}.
One finds that $R_{\eta}$
is approximately energy independent with a value of
$\sim6.5$
in the energy range of $16 - 109$ MeV
signifying a strong isovector exchange
contribution to the $\eta$ production mechanism.

The existing predictions for the $R_{\eta^\prime}$ differ drastically
depending on the model~\cite{cao-prc78,kampfer-ep}.
Yet, the ratio $R_{\eta^\prime}$ has not been measured to date, and only recently 
an upper limit has been determined~\cite{joanna} but this can be
significantly improved with the WASA-at-COSY experiment.

\section{Spin observables}
In the last decade a vast set of the unpolarised observables
has been established at the facilities CELSIUS, COSY and SATURNE
for the $\eta$ meson production in the collision of nucleons~\cite{hab,Moskal:2002jm}.
The data comprise in principle a lot of interesting information
concerning the production mechanism and the $\eta$-nucleon interaction.
These, however, could have not been derived unambiguously due to lack
of the knowledge about the relative contributions from the partial
waves involved.
One of the interesting unsolved problem as regards the $pp\to pp\eta$ reaction is the difficulty
in reproducing the $pp$ invariant mass distributions~\cite{abdel, prc69, christian, pm10}.
Calculations which include $NN$ FSI and $N\eta$ FSI
do not match existing data \cite{prc69}.
To explain the unexpected shape of the distribution,
 possibility of higher partial-waves is considered.
Taking into account a $P$-wave contribution one could
reproduce the $pp$ invariant mass distribution 
but not the close to threshold cross section dependencies \cite{Nakayama:2003pw}.
To solve this discrepancy, a 
$D_{13}$ resonance has been included~\cite{kanzoMENU} 
in the calculations. However, the data collected so far are insufficient for the unambiguous extraction of the $S$-wave or $P$-wave contributions.
Up to now there are only three measurements of the analysing power for the $\vec{p}p\to pp\eta$
reaction which have  been performed  with  low statistics and
the determined value of analysing power is essentially consistent with zero~\cite{rafalprl,pwinter,disto}
within large error bars of about $\pm$0.15.

Therefore recently, 
the azimuthally symmetric WASA detector and the polarised proton beam of COSY,
have been used~\cite{spin2010} to collect a high statistics sample of
$\vec{p}p\to pp\eta$ reactions in order to determine  
the analysing power as a function of the invariant mass spectra of the two particle subsystems 
and subsequently
to perform the
partial wave decomposition  with an accuracy by far better than resulting from 
measurements of the distributions of the spin averaged
cross sections.
The expected result should shed a light on the still not explained  origin of  structures
in the invariant mass distributions observed independently by the TOF~\cite{abdel},
COSY-11~\cite{prc69,pm10}, and CELSIUS/WASA~\cite{christian} collaborations.
It is worth to stress that similar shapes of the invariant mass distributions have
been also observed recently in the case of the $\eta^{\prime}$ meson~\cite{c11klaja}.
In both the $\eta$ and the $\eta^{\prime}$ case the intricate structure remains so far unexplained.

\section{Acknowledgements}
The author is grateful to the KLOE-2 and WASA-at-COSY Colleagues 
for the kind help in the preparation
of the talk 
and appreciates corrections of the manuscript by Fabio Bossi, Andrzej Kup{\'s}{\'c} and Steven Bass.
The author acknowledges also support by  Polish Ministery of Science and
Higher Education through the Grant No. 0469/B/H03/2009/37,
by the INFN, by the FFE
grants from the Research Center J\"{u}lich,
by the MPD programme of Foundation for Polish Science
through structural funds of the European Union,
by the FP7 Research Infrastructure \emph{HadronPhysics2}
(INFRA--2008--227431) and by the PrimeNet.

\end{document}